\documentclass[runningheads,a4paper]{llncs}

\usepackage{amssymb}
\usepackage{graphicx}
\usepackage{todonotes}
\usepackage{caption}
\usepackage{xcolor}
\usepackage{tabularx}
\usepackage{cite}
\usepackage{pbox}
\usepackage{listings}
\usepackage{url}
\urldef{\mailsa}\path|{alfred.hofmann, ursula.barth, ingrid.haas, frank.holzwarth,|
\urldef{\mailsb}\path|anna.kramer, leonie.kunz, christine.reiss, nicole.sator,|
\urldef{\mailsc}\path|erika.siebert-cole, peter.strasser, lncs}@springer.com|

\begin{document}

\mainmatter  

\title{Securing Application with Software Partitioning: A case study using SGX}

\titlerunning{Securing Application with Software Partitioning: A case study on SGX}
\authorrunning{A. Atamli, and A. Martin}
%
%
\author{Ahmad Atamli-Reineh\inst{1}, Andrew Martin\inst{1}}
\institute{Department of Computer Science, University of Oxford \email{[firstname.lastname]@cs.ox.ac.uk}}

%


%
%

\maketitle

\begin{abstract}

Application size and complexity are the underlying cause of numerous security vulnerabilities in code. 
In order to mitigate the risks arising from such vulnerabilities, various techniques have been proposed to isolate the execution of sensitive code from the rest of the application and from other software on the platform (e.g. the operating system). 
However, even with these partitioning techniques, it is not immediately clear exactly \emph{how} they can and should be used to partition applications.
What overall partitioning scheme should be followed; what granularity of the partitions should be.
To some extent, this is dependent on the capabilities and performance of the partitioning technology in use.
For this work, we focus on the upcoming Intel® Software Guard Extensions (SGX) technology as the state-of-the-art in this field. 
SGX provides a trusted execution environment, called an \emph{enclave}, that protects the integrity of the code and the confidentiality of the data inside it from other software, including the operating system. 
We present a novel framework consisting of four possible schemes under which an application can be partitioned.
These schemes range from coarse-grained partitioning, in which the full application is included in a single enclave, through ultra-fine partitioning, in which each application secret is protected in an individual enclave.
We explain the specific security benefits provided by each of the partitioning schemes and discuss how the performance of the application would be affected.
To compare the different partitioning schemes, we have partitioned OpenSSL using four different schemes. We discuss SGX properties together with the implications of our design choices in this paper.

\end{abstract}

\section{Introduction}
Applications have grown tremendously in functionality and size. This growth in sensitive applications and libraries such as Apache and OpenSSL has long ago surpassed the feasible limit for assurance techniques such as formal verification to verify the correctness of the code, and numerous factors have rendered manual review equally insufficient for that task. Accompanying the growth of the code in these applications, more classes of vulnerabilities have been identified, such as stealing secrets and modifying sensitive code\cite{Misra2003}\cite{one1996smashing}. An example that demonstrates this was the HeartBleed bug in the OpenSSL library where an attacker was able to obtain sensitive information including user names and passwords, credentials, and sensitive keys from remote servers\cite{Sullivan2014}. 

Much research has considered the design of systems based on well-known operating systems and hardware components to protect sensitive code. Many of these systems leverage virtualisation and trusted computing to isolate the execution of the entire application \cite{england2003trusted, chen2008overshadow, martignoni2012cloud, garfinkel2003terra, ta2006splitting, Paverd2012, li2014minibox, hofmann2013inktag,atamli2014threat}. However, many applications have thousands lines of code which makes it hard to gain assurance that no vulnerability exists in the code. Moreover, when virtualisation is used to provide isolation between different executions, there are many trust assumptions that make these systems limited in their security properties. For example, the Virtual Machine Monitor (VMM) or the code providing isolation needs to be trusted, loading the Trusted Computing Base (TCB) with thousands lines of code. 
The TCB is defined by the size of code that runs inside the same environment such as an isolated environment. 
The isolation of a software partition protects the data and the execution from external code, e.g. the OS and applications running in the same system. It follows that software partitioning of the application into several trusted partitions and untrusted partition, is expected to produce smaller partitions of code when considering the whole application as one partition. 
The latter, when partitioning to smaller chunks is feasible, may allow to formally verify the partition, which is protected by an isolated environment from external code and vulnerabilities such as vulnerabilities in other partitions of the same application.

Other systems\cite{McCune:2008:FEI:1357010.1352625,McCune5504713,azab2011sice,sahita2009protecting,yee2009native,dewan2008hypervisor,singaravelu2006reducing,cheng2013appshield}  provide isolation for the execution of a sensitive code without defining the portion of the application running on the trusted space, the granularity of these approaches to port sensitive code, or the feasibility to port small code such as merely few methods of an existing library. For instance, the TrustVisor\cite{McCune5504713} authors appreciate the complexity of porting security sensitive code in trusted environment. Porting security sensitive code is straightforward if the program is privilege-separated and modular. However, it is a greatest challenge in complex applications such as Apache + OpenSSL \cite{McCune5504713}.

To overcome the above mentioned shortcomings, processor extensions have been proposed in several pieces of research \cite{ARM,McCune:2008:LYG:1353534.1346285} to protect software execution and reduce the TCB. 
Protecting the code execution of the TCB is achieved with \emph{Trusted Execution Environment} (TEE) in hardware, which prevents external software from tampering with the execution, or modifying an existing code/data. 
Intel has also proposed security extensions to Intel® Architecture called Intel® Software Guard Extensions (Intel® SGX)\cite{Hoekstra2013},extensions that enable provisioning of sensitive data within applications. 
These extensions allow an application to instantiate a protected container to ensure the confidentiality and integrity of the data even in the presence of malware, while also relying on hardware to prevent external access to the container's memory area.
The protected container protects the inner code/data from external software, even privileged one, and is referred to as an \emph{enclave}.

Generally, the code and data are freely available for inspection and analysis prior to loading them into the enclave. Once loaded into the enclave and measured, they become protected against external software access. 
In order to store data outside the enclave's boundary, e.g. on the disk, the application can request from the enclave to seal the data beforehand. 
Furthermore, the platform key, which is used to encrypt the data, ties the data to the platform and can be used to report platform identity to remote parties. Overall, these capabilities extend the ability of enterprises and personnel to design secure applications by relying strongly on hardware instead of traditional software techniques. The aforementioned hardware provides another layer of protection against exploits of vulnerabilities missed by the tools verifying the correctness of the code or in manual reviews.  


However, even though many technologies are available, it is not necessarily obvious exactly \emph{how} they can and should be used to partition applications.
For some simple cases, the choice of partitioning scheme might indeed be obvious, but as applications increase in size and complexity, the number of possible partitioning schemes increases and the choice of the optimal approach becomes a very important non-trivial consideration.
From a technical perspective, partitioning schemes vary in terms of the security guarantees they provide and their impact on the performance of the application.
The choice of partitioning scheme has also other indirect implications, such as the effort for the application developers or software maintainers, but these are beyond the scope of this paper.

In this paper we investigate different software partitioning schemes using protected container, a TEE, to protect secrets from vulnerabilities in applications. Each scheme defines a different TCB size in each partition, which has immediate consequences on the economics of the TCB assurance process, in particular, its relation to the number of undetected vulnerabilities. 


As a rule, we isolate software partitions as defined in each scheme, and use an enclave to protect its execution and data from access by untrusted code. 
Previous research \cite{Hoekstra2013}addressed the threat model and components of SGX; our paper explores the use of hardware primitives, such as those offered by SGX, to provide secure design of applications through partitioning to keep the confidentiality and integrity of application's data. We implement two of the four partitioning schemes using SGX and test their ability to protect the system against an exploit of the \textit{HeartBleed} bug.

Our main contributions are:
\begin{itemize}
\item Proposing framework for different software partitioning schemes of an application.
\item Investigation of different software-partitioning schemes using SGX, with an empirical focus.
\item Proposing and investigating an evaluation matrix for partitioning schemes. 

\end{itemize}

The paper is divided into seven sections. Section 2 provides a brief background on SGX and some of its instructions and features. Section 3 discusses the rationale of this paper, objectives, and adversary model. In section 4 we demonstrate the rationale behind software partitioning and several partitioning schemes. 
Section 5 presents a real-world case study partitioned based on our proposed schemes, with security and efficiency evaluation of each scheme. Section 6 discusses related work, and finally section 7 concludes the paper. 
\section{Background}
\subsection{Isolation Mechanisms}
In this section we list different mechanisms used for isolation, and briefly list examples of systems that make use of such mechanisms. 
\subsubsection{Software-Enforced Isolation}
There are several ways to create separation between partitions. The most common approach used in software is using privileged code such as an OS or Virtual Machine Monitor(VMM) that enforces access control semantics\cite{McCune5504713}. A VMM will typically use hardware assistance for virtualisation, however the access control is enforced by software using meta-data of a memory address table. In contemporary operating systems the OS enforces access control between processes. Each process has its own code and data in memory, and the OS prevents one process from accessing another process space, that includes memory addresses and code. 

\subsubsection{Hardware-Enforced Isolation}
In order to isolate a partition from the rest of the system, hardware primitives have been proposed to provide TEE\cite{ARM,McCune:2008:FEI:1357010.1352625}. The TEE isolates the code execution from the rest of the system in hardware and enforces memory access semantics between the code running in the TEE. We refer to the code in the TEE as trusted code, and the code of the rest of the system as untrusted code. Arm TZ allows switching to a TEE from the untrusted space on TEE instruction invocation: the hardware moves the processor to TEE mode where data and code are separated from the rest of the system. 



\subsection{Software Guard Extensions (SGX)}
An overview of the SGX protection model \cite{McKeen2013} was given by Mckeen et al. In their paper they present the core of this  technology, the extensions that enable instantiating a protected container, describe the SGX instruction set, security model, threat model, and the hardware component on which this technology is based. In this section we give the background on SGX and its protection capabilities that is relevant to this work.

\begin{itemize}
\item \textbf{Enclave} - Intel SGX provides hardware features that creates a form of user-level TEE. The enclave is an isolated region of code and data within an application's address space. Data within an enclave can be accessed only with code within the same enclave. The enclave is able to protect its data using Enclave Page Cache (EPC); a secure storage used by the processor to store pages when they are part of an executing enclave. The EPC is built from chunks of 4KB pages; aligned on a 4KB boundary and each page has security attributes in the Enclave Page Cache Map (EPCM), an internal micro-architecture structure that is not accessible by software. It tracks the content of each EPC page, and enforces access control for accessing the pages.
\item \textbf{Measurement} - a cryptographic hash of the code and data residing in an enclave at the time of initialisation. The measurement is used to verify that the loaded enclave is what the enclave claims it is. 
\end{itemize}


\subsubsection{SGX Enclave Instructions and Protection Rings}
The enclave instructions available with SGX are divided under two protection rings; ring 0 and ring 3\cite{schroeder1972hardware}. The allowed set of instructions is determined according to the privilege level of the executing software. For the most part, ring 0 instructions; ECREATE, EADD, and EINIT are used for EPC management thus executed by privileged software such as OS and VMM. While ring 3 instructions e.g. EENTER, EEXIT, EGETKEY, EREPORT, and ERESUME are used by the user-space software to execute functionality within or between enclaves.  

\subsubsection{Enclave Life Cycle}
In order to provide strong security features, managing an enclave is done in hardware through enclave build instructions. To create an enclave, ECREATE instruction is used. It builds the enclave and sets base and range addresses. Once an enclave is created, EADD is used to add 4KB protected pages of data and code. This is followed by measuring the enclave's content using EEXTEND to protect the integrity of the data within the enclave. 
To elaborate on the latter, adding and measuring the enclave's pages are done by software prior to EINIT instruction. Once called, it finalises the measurement of the enclave and establishes an enclave identity. Executing within an enclave prior to this instruction is not allowed. On success of EINIT, entry to the enclave is enabled and permitted to run on the processor in privileged mode called \textit{enclave mode}. 

In order to enter and exit the enclave under program control, EENTER and EEXIT are used respectively. On enclave entry, the cached addresses are flushed, including addresses that overlap with the addresses used by the enclave to ensure the protection of the memory accesses within the enclave. Similarly, on enclave exit any cached addresses referring to the protected space in an enclave are cleared. The purpose of this is to prevent external software from using the cached addresses to access the enclave's protected memory. 

\subsubsection{Asynchronous Exit and Resuming Execution}
Exiting the enclave asynchronously occurs due to events such as exceptions and interrupts in which the processor handles such events by invoking the internal routine Asynchronous Exit (AEX). 
The AEX saves the registers used by the enclave which are consequently cleared to prevent leaking secrets. In particular, one saved address to be stored is the location of the returning address, also called the faulting address, where the execution resumes on the resuming enclave's execution. While saving the enclave's state is essential for resuming the enclave's execution, equally important is clearing the data used by the enclave to prevent secret exposure. Once AEX finishes execution, the processor exits enclave mode and goes back to normal mode where every instruction is treated as an external instruction.

On the other hand, the ERESUME instruction restores the enclave's state and gives back control to the enclave from the point it was interrupted. It is important to mention that the event whom the AEX was called upon may be triggered again in case of failure when the event is an exception or faults within the enclave.


\section{Objectives and Adversary Model}
\subsection{Security Objectives}
Applications consist of data, e.g. keys, passwords, and code of third-party libraries such as OpenSSL.
Protecting secrets is a major priority; an application would like to keep the confidentiality and integrity of these secrets, and the integrity of the code executing using these secrets. The exposure of one element is enough to compromise the entire system. Furthermore, sensitive parts of an application constitute a small fragment of the code as a whole in most applications. Thus, isolating the data storage and execution of sensitive parts from the rest can decrease the impact of vulnerabilities. 

Our security objective is to keep the \emph{Confidentiality} even in the presence of malware (including malware running within the privileged operating system), and reduce the impact of vulnerabilities in code. 
It has been shown that hardware-assisted partitioning technology, such as Intel SGX, can be used to achieve this\cite{Hoekstra2013,baumann2014shielding}. 

The enclave keeps the confidentiality of the data by encrypting its content when leaving the processor in enclave mode e.g. in memory. Our objective is to protect secrets such as passwords, keys, and sensitive code from vulnerabilities in applications. One approach to achieving this when considering a trusted OS is to use a different process for each partition,relying on the OS to enforce memory access control semantics between processes. However, we assume untrustworthy OS, an OS that might have vulnerability or malware, thus, using the processes is not an option. To elaborate on the latter, we do not consider an OS that is untrustworthy as a result of an adversary booting malicious OS. We assume that the OS is coming from trusted source but may have vulnerabilities or malware which may risk the exposure of secrets in applications.
 
It is important to note that using systems with one TEE such as ARM TrustZone\cite{ARM}, and Flicker\cite{McCune:2008:FEI:1357010.1352625} does not scale in flexibility for partitioning applications. These systems address how to isolate trusted code from untrusted code using one TEE, and managing the TEE for different partitions requires intervention of software and not hardware. On the other hand,  SGX does allow instantiating of many containers using hardware operations, thus, it is well suited for our partitioning schemes and in evaluating the security of each scheme.

\subsection{Adversary Model}

In this paper, we consider an adversary with the capabilities to insert malware into the system, read the memory, and manipulate the OS including booting another OS. An adversary aims to exploit vulnerabilities in application code who may be able to obtain secrets or cause malfunction, which eventually may lead to exposure or modification of sensitive data. The adversary may have knowledge of the software running, but does not have physical access to the system's CPU and physical parts of the platform including memory controller or the buses interconnecting between platform components\cite{Zhang:2012:CSC:2382196.2382230,fan2010state}.
The adversary may be an insider with a limited physical access to the system, or a remote adversary. 
We do not aim to protect against attacks such as denial of service or side channel attacks.

\section{Application Software Partitioning}
\label{ASP}
We are proposing a partitioning scheme framework and that will be illustrated and explained with a concrete example of OpenSSL. However, the approach taken here is applicable to all types of applications that protect secret data.
In the trusted part we would like to port sensitive functions and data such as hashing functions, random number generator, certificates, keys and passwords. The untrusted code will be located out of the TEE with the ability to call protected functions to be executed in TEE. While the untrusted code may be able to request for encryption and decryption services from the trusted code, it is unable to read/write the keys and the cryptographic functions that reside within a TEE to provide these services. The untrusted code may merely call the interface TEE functions for execution. The trusted part is considered as a \textit{Black Box} to the untrusted part, thus, protecting the confidentiality and integrity of the code and data.

The application must be partitioned into several parts by identifying the sensitive partitions that require isolation from other parts of the application. The design guideline is to keep a sensitive partition minimal and within feasibility borders to allow formal verification of the code. While the TEE can protect its execution and secrets from external vulnerabilities, it does not protect against badly written code with flaws. Thus, a partition with small code is a corner stone for designing a secure application and has been long advocated by Saltzer and Schroeder\cite{saltzer1975protection}. However, it is important to bear in mind the efficiency of the execution when partitioning the code. A partition scheme that substantially impairs system efficiency will often be unfeasible regardless of its security characteristics.

\subsection{Partitioning Schemes}

In this section, we describe several possible partitions schemes. We start with basic partitioning configuration and develop it further as a function of the TCB and number of enclaves that yield different partitioning schemes. These schemes may differ in their ability to protect the confidentiality of the data, which we will be investigating in more details in section 5. 

Initially, we started by defining a partitioning scheme that considers two guidelines: 1) the number of available enclaves 2) the TCB size inside each enclave. In scheme~1, we started with the most basic configuration, one enclave and without any limitation on the size of the TCB inside that enclave. Our aim is clear and simple; to protect the secrets as described in detail in section 3 from the rest of the code. In scheme~2, we chose to increase the number of enclaves by one, two enclaves with a reduction in the TCB as explained in \ref{ref:schem2},  which led us to scheme~3. In Scheme~2, the size of the TCB inside each enclave is reduced to an optimal level. However, accounts/connections/users have to use the only available two enclaves, thus, no separation between the different accounts/connections/users. In scheme~3, we built on scheme~2 and adopted a similar TCB inside an enclave but with open approach toward the number of enclaves that isolate between different accounts/connections/users. Scheme~3 proved to be very complex both for security and implementation. For instance, a trusted channel is needed between every two enclaves that wished to talk to each other, thus, with the adopted open approach in scheme~3 many trusted channels are needed. Also, with this approach every piece of code inside an enclave needed to be duplicated for full separation between the accounts/connections/users. Hence, we identified a potential implementation and performance issues prior to evaluating the approach. It follows, in scheme~4 we took scheme~3 and optimised it by considering reducing the number of enclaves, TCB, and duplication of code.

\subsubsection{Scheme 1 - Whole Application}
In this scheme we choose to put part of an application such as a library inside one enclave. The residents of the enclave which may be code and data, include all secrets such as keys (e.g. private key, storage key, session key), passwords, credentials, and the code. 

\subsubsection{Scheme 2 - All Secrets}\label{ref:schem2}
In this scheme we apply smaller granularity compared to scheme~1. 
We use two enclaves, we divide the code in two partitions, based on the frequency of accessing the code and port the code that generates secrets and has high frequency for accessing the secrets. The rationale is to opt-out the code that does not have high frequency of accessing the secrets which will result in reduction of code's  lines number, hence, reduction of the TCB. However, it is important to mention that an application with different users has all its users' secrets within the same enclave. Thus, it is the responsibility of the software running inside the same enclave to enforce isolation between users' data.

\subsubsection{Scheme 3 - Separate Secret}
Scheme 3 is smaller in granularity compared to the previous two schemes. 
We use multiple enclaves to secure the secretes. Each enclave contains one secret such that each key resides in a separate enclave. For example code using the session key lies in one enclave and code using the private key lies in another enclave. We use multiple enclaves per account/user/connection, where each enclave contains the secrets generation relevant code and its relevant key, and one enclave for the code that has high frequency of accessing the code after generation.
\subsubsection{Scheme 4 - Hybrid}
In this scheme we apply smaller granularity than Scheme~1 and Scheme~2 but less than Scheme~3. 
We use multiple enclaves to protect Application's secretes. Each account/user/connection has a separate enclave. One enclave per account/user/connection, that includes keys (e.g. private key, session key), generation code, and functions with high frequency for accessing the secrets. For example, an application with multiple users, each user's secrets reside within the same enclave. However, in order to reduce the number of enclaves used, we use an enclave that contains code but not secrets to give services to all accounts. When a secret is needed, it's sent to another enclave which is assumed not to store any data. This scheme is similar to scheme 2 in the definition of the TCB residing inside an enclave, however, while scheme 2 has all secrets of all users/connections/account in an enclave, scheme 4 isolates between users/connections/account by having enclave for each. On the other hand, scheme 4 is similar to scheme 3 in the way it isolates the secrets of each users/connections/account.

\subsection{Partitioning using SGX}
The application uses SGX to protect the execution of sensitive partitions by porting different sensitive partition into different enclaves. The number of TCBs is the influential factor for the number of partitions constructed prior to running the application, and during the run time, SGX enforces access between these partitions.
It is important to mention that porting the code to run in trusted space is not the only action required when partitioning the code, the same ported code should be able to handle I/O operations and external operations and exit enclave mode when necessary.
The interface to the enclave is limited and the creation process requires the intervention of privileged software that runs in ring~0, e.g. SGX driver. As a rule, the privileged software creates an enclave using ECREATE, adds, and measures the code of the desired partition. It uses EADD and EEXTEND respectively to perform the latter, which is then followed by EINIT to finalise the creation process, and entering the enclave by the same application that created it.
In order to enter an enclave, the application uses synchronous entry instruction EENTER to switch the processor to enclave mode and to execute the relevant call. 

As an essential part of the design, I/O operations are excluded from the enclave since they require the intervention of the OS, thus, when I/O operation is required, synchronous exit (EEXIT) is called to switch the processor to normal mode to handle the requested external operation. In a similar way the OS interrupts are handled through \textit{Asynchronous Exit and Resuming Execution instructions}. Once done, the trusted part resumes by re-entering the enclave with ERESUME. 

Once the enclave finishes execution it exits the enclave mode using EEXIT and the processor returns to normal mode of execution. The life cycle of the enclave and its content can be terminated by the application using privileged software; the privileged software tears down the pages inside the enclave (EREMOVE) and removes all the meta-data associated with it.

\section{Security and Efficiency Evaluation}
\label{sec:Security_Evaluation}
We use a \emph{MiniServer + OpenSSL} library to examine several software partitioning schemes. The MiniServer is a web server that serves multiple clients and provides authentication, and secure communication channel. The MiniServer runs on Linux and uses merely minimal code to establish secure connections with clients. Furthermore, it uses the OpenSSL library for establishing secure connection between the server and the client\cite{OpenSSL}.  

To meet our objectives we choose to consider two main components in the SSL protocol: the handshake protocol and the data exchange. During the handshake, the client and the server generate keys which are unique for each connection session. The session defines a set of cryptographic security parameters which can be shared among multiple connections. For the most part, the handshake protocol allows the server and the client to authenticate each other and to negotiate a cryptographic suit. The handshake protocol consists of several messages exchanged between a client and a server prior to establishing a secure channel.  It is followed by the second part of the protocol execution in which data is exchanged between client and server. 

In order to evaluate the security and efficiency of the proposed schemes we consider partitioning the OpenSSL library 1.0.2-beta1. On the security side we investigate: 1) the ability of a scheme to protect against vulnerabilities in code such as the HeartBleed vulnerability; 2) the number of trusted channels required between partitions; and 3) the size of the TCB. Our primary reason for considering these evaluation items is their impact on the attack surface. For example, the size of the TCB has a direct impact on the number of vulnerabilities in code. Also, an application with various enclaves requires trusted channels for communicating between these enclaves, thus increasing the complexity of the system and expanding the attacks surface since there are more components to protect.
On the efficiency side, we consider the number of enclaves, the number of entries to these enclaves, and the size of each enclave. Moreover, context switching is required when moving in to and out of the enclave, introducing an overhead that increases with the number of enclaves and entries to these enclaves.
We evaluate the security and efficiency of the proposed partitioning schemes from section~\ref{ASP} and present the calculated results in table \ref{table1}.

\subsection{Case Study}
In this section we use the OpenSSL library to examine the proposed software partitioning schemes. In particular, we choose a vulnerability from the buffer over-read class of attacks, the \textit{HeartBleed} vulnerability~\cite{HeartBleed}, to evaluate each scheme. The aforementioned vulnerability will demonstrate the ability of each scheme to meet our objective of protecting the private and session keys. While, a straightforward solution is to fix the vulnerability when found, our proposed method of isolating software partitions from each other aims to counter the over-read class of attacks when a vulnerability is missed during the verification process. 


The vulnerability known as HeartBleed results from missing bounds check in the heart beat extension which is a `keep-alive' mechanism between two endpoints to keep the connection alive. The latter was classified as a buffer over-read vulnerability and it allows more data to be read than was initially negotiated between the client and server, thus revealing secrets and sensitive data. The sensitive data is not limited to secret keys used within the OpenSSL library, but also includes user names and passwords of the application that happen to be in the requested memory space. For the most part, applications rely on privileged software such as the OS to prevent external access to an application space. However, in the presence of vulnerability in an application such as in a third-party library, the OS does not play any part in protecting the data of the entire application, specifically, data that is generated by the application but not used by the imported third-party library.

\subsection{First Scheme - Whole Application as One Partition}
In the first scheme the entire SSL library resides in a single enclave and includes the heart beat code. The code within an enclave has memory access to every memory address inside the same enclave, thus when a client requests more data than it has sent, the heart beat code is still able to extract the requested length, notwithstanding its content e.g. session and private keys, and send it back to the client. Moreover, data from the application using OpenSSL, such as user-names and passwords, can be extracted when residing in adjacent memory addresses to the requested data. Hence, the rest of the application is vulnerable to secrets exposure. 

Using TEE does not protect against vulnerabilities in the code. While the data is protected with encryption from external software when it resides in the memory, it is not protected from vulnerabilities that reside in the enclave. To illustrate this using the HeartBleed example, the heart beat code resides within an enclave, thus it is part of the same TCB that contains the secret keys and functions used during the SSL session. As a result, the security properties provided by the enclave are transparent to the contained software, and accessing secrets from an inner function, such as the heart beat code, can be achieved without the enclave's interference.

Scheme 1 uses one enclave and thus doesn't require any trusted channels. However, the big drawback is the large size of TCB that includes the buffer over-read vulnerability, which in return it doesn't protect the confidentiality of secrets upon implementation. 

\subsection{Second Scheme - All Secrets}
In the second scheme we used two enclaves to isolate part of the OpenSSL library including the handshake protocol, private key, session key, and data exchange. We partition the code such that only key handling the code (both session and private) are inside the enclave, but heartbleed code is outside that enclave.  

Scheme 2 protects against exploitation of the HeartBleed vulnerability since the heart-beat code can not access the session key which is encrypted in memory as part of an enclave. The TCB is smaller than that of scheme 1. However, other secrets of the application, such as the user-names and passwords of the server which are not part of the enclave, are not protected. Also, one might question the security of having all the session keys within the same enclave used by the same code. To state the obvious, mutual exclusion between the different sessions is not achieved with this scheme. 

\subsection{Third Scheme - Separate Secrets}
In scheme 3 each connection has two enclaves, one for the handshake protocol and session key, and one for the data exchange. To elaborate on the latter, since each connection has two enclaves, it's obvious that some duplication of code is inevitable. Nonetheless, the private key resides in a different enclave and can be used by other enclaves that require access to it.

In scheme 3 isolating each secret in a different enclave protects against code vulnerabilities, such as HeartBleed, compromising the confidentiality or integrity of the session key or private key. The TCB in each of the enclaves is significantly smaller than in scheme 1 . However, this approach brings with it other challenges: In order to prevent malicious software from exploiting the different enclaves, a trusted channel must be established between the different enclaves to assure secure communication and execution of the partitions combined. The latter may impair the execution efficiency in favour of isolating connections. However, more detailed empirical work is needed to examine this, which is beyond the scope of this paper.

\subsection{Fourth Scheme - Hybrid Software Partitioning}
In this approach we considered a hybrid partitioning of the code, which is a combination of the aforementioned schemes. The main code resides in the untrusted space and only a part of the code and data resides in the enclave. The heart beat code resides in the untrusted space of the application and is thus unable to access the secrets within the enclave. The heart beat code could reside in a separate enclave if need be. The main focus of our design is on partitioning the application in such a way that sensitive partitions with secrets are isolated from other unrelated partitions. In scheme 4, the TCB is smaller than in schemes 1 and isolation between the sessions is achieved. However, TCB is not as small as in scheme 3. The advantage of scheme 4 over scheme 3 is a reduction in the number of enclaves. The number of trusted channels required between different enclaves is smaller, which results in less overhead in the system and the trusted channel being a target for adversaries.
To test this framework, we implemented the hybrid approach using SGX - a combination that proved to be resilient to read-overflow vulnerabilities such as HeartBleed. 
In addition, with this scheme the size of the TCB inside the enclave proved to be much smaller than scheme 1.  In table \ref{table1} we summarise the analysis of the 4 different partition schemes discussed.

\begin{minipage}[c]{\linewidth}

\captionof{table}{Comparison between the 4 schemes} \label{table1} 

\bigskip
\begin{tabular}{ p{7cm} | c | c | c | c |}

    &\parbox[t]{10mm}{\centering \rotatebox[origin=c]{90}{  Whole Application }}  &\parbox[t]{10mm}{\centering \rotatebox[origin=c]{90}{All Secrets}}  &\parbox[t]{10mm}{\centering \rotatebox[origin=c]{90}{Separate Secret}}  &\parbox[t]{10mm}{\centering \rotatebox[origin=c]{90}{Hybrid}}\\
 \hline
 \parbox{20cm}{\smallskip Number of Enclaves \\ \footnotesize{(10 Connections)} \smallskip}  &1  &2  &21 &11 \\
 \hline
 \parbox{20cm}{\smallskip Trusted Channels between Enclaves  \\\footnotesize{(One connection)} \smallskip} &0  &0  &3 &2 \\
 \hline
 \parbox{20cm}{\smallskip TCB in enclave  \smallskip} &L  &S  &S &S  \\
 \hline
 \parbox{20cm}{\smallskip Duplication of Code  \smallskip} &No  &No  &Yes &Yes \\
 \hline
 \parbox{20cm}{\smallskip Capacity Used \smallskip}  &M  &S  &L &M - L \\
 \hline
\end{tabular}
\newline
\newline
\centering{Size Scale : L - Large,  M - Medium , S - Small }
\end{minipage}

\section{Related Work}
In the last decade the topic of executing sensitive code in isolated and trusted environment has caught the attention of many researchers. McCune et al. presented Flicker\cite{McCune:2008:FEI:1357010.1352625}- an infrastructure for code execution in isolated and trusted environment. In their work they rely merely on 250 lines of code in the TCB to provide strong isolation. For the most part, they appreciate that 250 lines of code is a tiny code, therefore formal assurance of its execution is more trusted as a result of the feasibility to verify the code. Nonetheless, an application running in an isolated execution environment can be thousands of line of code and isolation between several parts in the application space is essential to prevent exploits by unfortunate vulnerabilities. The same group presented TrustVisor\cite{McCune5504713} a pointed purpose hypervisor that provides code and data integrity and secrecy for sensitive portions of an application. TrustVisor provides application developers with a strong secure environment for code execution and data storage on untrusted platforms. Moreover, they argue that small TCB code is easier to be formally verified, thus, it is more trusted when executing in TEE. Another research effort that takes a similar approach is that of Singaravelu et al.~\cite{Singaravelu:2006:RTC:1218063.1217951} where they showed that reducing TCB complexity can result in enhancing the security of the sensitive part of the application. The sensitive part is executed in a process called AppCore while the rest of the application is executed on a virtualised untrusted operating system. This approach is supported by three real world case-study applications.

In \cite{Strackx:2012:FSH:2382196.2382200} Strackx proposed Fides: a security architecture that consists of two parts: a run-time security architecture and a compiler. The run-time security architecture is based on memory access control to protect applications. The modules are divided into a private section, where sensitive data is protected and accessed by the relevant module through limited interface, and a public section that contains the module's code. The second part is the compiler which is responsible for compiling standard C code into protected modules. In another work \cite{Cheng:2011:DFP:2041225.2041242}, Cheng et al. presented DriverGuard, a hypervisor protection mechanism to shield I/O flow from a malicious kernel. DriverGuard protects a tiny fraction of the code that is sensitive, such as biometric authentication. However, they assume secure boot-up and load-time attestation to ensure the hypervisor's security in the bootstrapping phase. 

In \cite{li2014minibox} Li et al. introduce MiniBox, a two way sandbox that isolates the memory space between OS protection modules and applications. Unlike most approaches it aims to protect the OS from untrusted applications, but also protects the applications from a malicious OS. In Minibox, the authors focus on the two-way Sandboxing and don't address the porting efforts for legacy code, and suffice by mentioning that the porting efforts are similar to the porting effort on NaCl~\cite{yee2009native}.

In \cite{Vasiliadis:2014:PUG:2660267.2660316} Vasiliadis et al. introduce PixelVault, a system that uses GPUs to secure cryptographic keys.  In PixelVault the private key is created inside the GPU and never leaves or leaks it even in the presence of malicious OS. However, this is limited to the private key since PixelVault can not use the GPU to secure keys negotiated at run-time such as the session key or key pairs. Thus, malicious software can act as a man in the middle.

Partitioning privileges between hardware and software is not a new paradigm~\cite{Stitt:2003:DHP:775832.775896}. Hardware/Software partitioning has shown improvement in performance, energy consumption, and optimised run-time. However, there hasn't been much work that addresses hardware and software partitioning from security point of view. 

Our approach differs in the granularity and feasibility of isolating sensitive code. Most approaches rely on software to isolate the execution of sensitive code from the rest of the system. These approaches face significant difficulties when partitioning the code into trusted and untrusted sections. While it is straightforward to isolate an entire application using SGX, it is still feasible for programmers to partition the code into trusted and untrusted sections even when the application is not modular or privilege-separated. Unlike some hardware-based isolation techniques, SGX enables concurrent execution of more than one secure enclave. This allows applications to use various different partitioning schemes to achive the required balance between security and performance.

\section{Conclusions and Future Work}
In order to protect the execution of sensitive code and data, it is desirable to use a trusted execution environment that does not include untrusted entities such as the OS. This can be achieved by keeping the TCB as small as possible and excluding irrelevant parts of the code. 
Fine-grained software partitioning of the code provides a good means of isolating different parts of the application and defining trust relationships between the partitions. Such an approach can protect the execution of a sensitive code from untrusted partitions when access is enforced properly. SGX proves to be a good candidate that keeps the OS out of the TCB and protects the execution of a partition from untrusted code using hardware. 
It is widely expected that the adoption of technologies like SGX will facilitate the design of secure applications and add another level of protection against various vulnerabilities in the code. In this paper we have proposed a framework that describes exactly \emph{how} these technologies could be used to achieve this. We have explored four possible partitioning schemes that differ in terms of security guarantees and performance. We have demonstrated how our schemes could be realized using SGX to secure the execution of low level sensitive code in the SSL library as a proof of concept to our claims.

Another key point is that although the TEE is an important and desirable security feature, it is not a silver bullet against vulnerabilities in code.
We demonstrate a logical use of TEE and the feasibility of different software partitioning schemes with SGX in merely one example: the OpenSSL library. 
In future work we plan to perform broader research on fine-grained software partitioning using SGX with different applications that includes bench-marking each of the schemes described above. 
Eventually, we intend to develop a methodology to help developers partition applications effectively using these new technologies in order to balance security with performance.

\raggedright 

\bibliography{SSL}

\begin{thebibliography}{10}

\bibitem{Misra2003}
Misra, S.C., Bhavsar, V.C.:
\newblock {Computational Science and Its Applications — ICCSA 2003}. Volume
  2667 of Lecture Notes in Computer Science.
\newblock Springer Berlin Heidelberg, Berlin, Heidelberg (June 2003)

\bibitem{one1996smashing}
One, A.:
\newblock Smashing the stack for fun and profit.
\newblock Phrack magazine \textbf{7}(49) (1996)  14--16

\bibitem{Sullivan2014}
Sullivan, N.:
\newblock {Staying ahead of OpenSSL vulnerabilities | CloudFlare Blog} (2014)

\bibitem{england2003trusted}
England, P., Lampson, B., Manferdelli, J., Peinado, M., Willman, B.:
\newblock A trusted open platform.
\newblock Computer \textbf{36}(7) (2003)  55--62

\bibitem{chen2008overshadow}
Chen, X., Garfinkel, T., Lewis, E.C., Subrahmanyam, P., Waldspurger, C.A.,
  Boneh, D., Dwoskin, J., Ports, D.R.:
\newblock Overshadow: a virtualization-based approach to retrofitting
  protection in commodity operating systems.
\newblock In: ACM SIGOPS Operating Systems Review. Volume~42., ACM (2008)
  2--13

\bibitem{martignoni2012cloud}
Martignoni, L., Poosankam, P., Zaharia, M., Han, J., McCamant, S., Song, D.,
  Paxson, V., Perrig, A., Shenker, S., Stoica, I.:
\newblock Cloud terminal: Secure access to sensitive applications from
  untrusted systems.
\newblock In: USENIX Annual Technical Conference. (2012)  165--182

\bibitem{garfinkel2003terra}
Garfinkel, T., Pfaff, B., Chow, J., Rosenblum, M., Boneh, D.:
\newblock Terra: A virtual machine-based platform for trusted computing.
\newblock In: ACM SIGOPS Operating Systems Review. Volume~37., ACM (2003)
  193--206

\bibitem{ta2006splitting}
Ta-Min, R., Litty, L., Lie, D.:
\newblock Splitting interfaces: Making trust between applications and operating
  systems configurable.
\newblock In: Proceedings of the 7th symposium on Operating systems design and
  implementation, USENIX Association (2006)  279--292

\bibitem{Paverd2012}
Paverd, A., Martin, A.:
\newblock Hardware security for device authentication in the smart grid.
\newblock In: First Open EIT ICT Labs Workshop on Smart Grid Security -
  SmartGridSec12, Berlin, Germany (2012)

\bibitem{li2014minibox}
Li, Y., McCune, J., Newsome, J., Perrig, A., Baker, B., Drewry, W.:
\newblock Minibox: A two-way sandbox for x86 native code.
\newblock In: 2014 USENIX Annual Technical Conference (USENIX ATC 14), USENIX
  Association (2014)

\bibitem{hofmann2013inktag}
Hofmann, O.S., Kim, S., Dunn, A.M., Lee, M.Z., Witchel, E.:
\newblock Inktag: secure applications on an untrusted operating system.
\newblock ACM SIGPLAN Notices \textbf{48}(4) (2013)  265--278

\bibitem{atamli2014threat}
Atamli, A.W., Martin, A.:
\newblock Threat-based security analysis for the internet of things.
\newblock In: Secure Internet of Things (SIoT), 2014 International Workshop on,
  IEEE (2014)  35--43

\bibitem{McCune:2008:FEI:1357010.1352625}
McCune, J.M., Parno, B.J., Perrig, A., Reiter, M.K., Isozaki, H.:
\newblock Flicker: An execution infrastructure for tcb minimization.
\newblock SIGOPS Oper. Syst. Rev. \textbf{42}(4) (April 2008)  315--328

\bibitem{McCune5504713}
McCune, J., Li, Y., Qu, N., Zhou, Z., Datta, A., Gligor, V., Perrig, A.:
\newblock Trustvisor: Efficient tcb reduction and attestation.
\newblock In: Security and Privacy (SP), 2010 IEEE Symposium on. (May 2010)
  143--158

\bibitem{azab2011sice}
Azab, A.M., Ning, P., Zhang, X.:
\newblock Sice: a hardware-level strongly isolated computing environment for
  x86 multi-core platforms.
\newblock In: Proceedings of the 18th ACM conference on Computer and
  communications security, ACM (2011)  375--388

\bibitem{sahita2009protecting}
Sahita, R., Warrier, U., Dewan, P.:
\newblock Protecting critical applications on mobile platforms.
\newblock Intel Technology Journal \textbf{13}(2) (2009)

\bibitem{yee2009native}
Yee, B., Sehr, D., Dardyk, G., Chen, J.B., Muth, R., Ormandy, T., Okasaka, S.,
  Narula, N., Fullagar, N.:
\newblock Native client: A sandbox for portable, untrusted x86 native code.
\newblock In: Security and Privacy, 2009 30th IEEE Symposium on, IEEE (2009)
  79--93

\bibitem{dewan2008hypervisor}
Dewan, P., Durham, D., Khosravi, H., Long, M., Nagabhushan, G.:
\newblock A hypervisor-based system for protecting software runtime memory and
  persistent storage.
\newblock In: Proceedings of the 2008 Spring simulation multiconference,
  Society for Computer Simulation International (2008)  828--835

\bibitem{singaravelu2006reducing}
Singaravelu, L., Pu, C., H{\"a}rtig, H., Helmuth, C.:
\newblock Reducing tcb complexity for security-sensitive applications: Three
  case studies.
\newblock In: ACM SIGOPS Operating Systems Review. Volume~40., ACM (2006)
  161--174

\bibitem{cheng2013appshield}
Cheng, Y., Ding, X., Deng, R.:
\newblock Appshield: Protecting applications against untrusted operating
  system.
\newblock Singaport Management University Technical Report, SMU-SIS-13
  \textbf{101} (2013)

\bibitem{ARM}
ARM:
\newblock {ARM TrustZone}

\bibitem{McCune:2008:LYG:1353534.1346285}
McCune, J.M., Parno, B., Perrig, A., Reiter, M.K., Seshadri, A.:
\newblock How low can you go?: Recommendations for hardware-supported minimal
  tcb code execution.
\newblock SIGARCH Comput. Archit. News \textbf{36}(1) (March 2008)  14--25

\bibitem{Hoekstra2013}
Hoekstra, M., Lal, R.:
\newblock {Using innovative instructions to create trustworthy software
  solutions}.
\newblock Proceedings of the \ldots (2013)

\bibitem{McKeen2013}
McKeen, F., Alexandrovich, I., Berenzon, A.:
\newblock {Innovative instructions and software model for isolated execution}.
\newblock HASP (2013)

\bibitem{schroeder1972hardware}
Schroeder, M.D., Saltzer, J.H.:
\newblock A hardware architecture for implementing protection rings.
\newblock Communications of the ACM \textbf{15}(3) (1972)  157--170

\bibitem{baumann2014shielding}
Baumann, A., Peinado, M., Hunt, G.:
\newblock Shielding applications from an untrusted cloud with haven.
\newblock In: USENIX Symposium on Operating Systems Design and Implementation
  (OSDI). (2014)

\bibitem{Zhang:2012:CSC:2382196.2382230}
Zhang, Y., Juels, A., Reiter, M.K., Ristenpart, T.:
\newblock Cross-vm side channels and their use to extract private keys.
\newblock In: Proceedings of the 2012 ACM Conference on Computer and
  Communications Security. CCS '12, New York, NY, USA, ACM (2012)  305--316

\bibitem{fan2010state}
Fan, J., Guo, X., De~Mulder, E., Schaumont, P., Preneel, B., Verbauwhede, I.:
\newblock State-of-the-art of secure ecc implementations: a survey on known
  side-channel attacks and countermeasures.
\newblock In: Hardware-Oriented Security and Trust (HOST), 2010 IEEE
  International Symposium on, IEEE (2010)  76--87

\bibitem{saltzer1975protection}
Saltzer, J.H., Schroeder, M.D.:
\newblock The protection of information in computer systems.
\newblock Proceedings of the IEEE \textbf{63}(9) (1975)  1278--1308

\bibitem{OpenSSL}
{OpenSSL Software Foundation}:
\newblock {OpenSSL Library Version 1.0.2a}

\bibitem{HeartBleed}
{ N. Mehta and Codenomicon}:
\newblock {The Heartbleed Bug}

\bibitem{Singaravelu:2006:RTC:1218063.1217951}
Singaravelu, L., Pu, C., H\"{a}rtig, H., Helmuth, C.:
\newblock Reducing tcb complexity for security-sensitive applications: Three
  case studies.
\newblock SIGOPS Oper. Syst. Rev. \textbf{40}(4) (April 2006)  161--174

\bibitem{Strackx:2012:FSH:2382196.2382200}
Strackx, R., Piessens, F.:
\newblock Fides: Selectively hardening software application components against
  kernel-level or process-level malware.
\newblock In: Proceedings of the 2012 ACM Conference on Computer and
  Communications Security. CCS '12, New York, NY, USA, ACM (2012)  2--13

\bibitem{Cheng:2011:DFP:2041225.2041242}
Cheng, Y., Ding, X., Deng, R.H.:
\newblock Driverguard: A fine-grained protection on i/o flows.
\newblock In: Proceedings of the 16th European Conference on Research in
  Computer Security. ESORICS'11, Berlin, Heidelberg, Springer-Verlag (2011)
  227--244

\bibitem{Vasiliadis:2014:PUG:2660267.2660316}
Vasiliadis, G., Athanasopoulos, E., Polychronakis, M., Ioannidis, S.:
\newblock Pixelvault: Using gpus for securing cryptographic operations.
\newblock In: Proceedings of the 2014 ACM SIGSAC Conference on Computer and
  Communications Security. CCS '14, New York, NY, USA, ACM (2014)  1131--1142

\bibitem{Stitt:2003:DHP:775832.775896}
Stitt, G., Lysecky, R., Vahid, F.:
\newblock Dynamic hardware/software partitioning: A first approach.
\newblock In: Proceedings of the 40th Annual Design Automation Conference. DAC
  '03, New York, NY, USA, ACM (2003)  250--255

\end{thebibliography}
\bibliographystyle{splncs}

\end{document}